\documentclass[final,5p]{elsarticle}

\usepackage{graphicx}
\usepackage{amssymb}
\usepackage[english]{babel}

\newcommand{\avga}{\langle \alpha \rangle}
\newcommand{\avgae}{\langle \alpha_E \rangle}

\title{Electronic transport on carbon nanotube networks: a multiscale computational approach}

\author{Luiz F. C. Pereira \fnref{label1} }
\ead{pereirlf@tcd.ie}
\author{M. S. Ferreira \corref{cor1} \fnref{label2} }
\ead{ferreirm@tcd.ie}
\address{School of Physics, Trinity College Dublin, Dublin 2, Ireland.}
\cortext[cor1]{Corresponding author}
\fntext[label1]{pereirlf@tcd.ie}
\fntext[label2]{ferreirm@tcd.ie}

\date{\today}

\biboptions{sort&compress}

\journal{Nano Communication Networks}

\begin{document}

\begin{frontmatter}

\begin{abstract}
Carbon nanotube networks are one of the candidate materials to function as malleable, transparent, conducting films, with the technologically promising application of being used as flexible electronic displays. Nanotubes disorderly distributed in a film offers many possible paths for charge carriers to travel across the entire system, but the theoretical description of how this charge transport occurs is rather challenging for involving a combination of intrinsic nanotube properties with network morphology aspects. Here we attempt to describe the transport properties of such films in two different length scales. Firstly, from a purely macroscopic point of view we carry out a geometrical analysis that shows how the network connectivity depends on the nanotube concentration and on their respective aspect ratio. Once this is done, we are able to calculate the resistivity of a heavily disordered networked film. Comparison with experiment offers us a way to infer about the junction resistance between neighbouring nanotubes. Furthermore, in order to guide the frantic search for high-conductivity films of nanotube networks, we turn to the microscopic scale where we have developed a computationally efficient way for calculating the ballistic transport across these networks. While the ballistic transport is probably not capable of describing the observed transport properties of these films, it is undoubtedly useful in establishing an upper value for their conductivity. This can serve as a guideline in how much room there is for improving the conductivity of such networks. 

\end{abstract}

\begin{keyword}
carbon nanotube networks \sep Green functions \sep electronic transport 
\end{keyword}

\end{frontmatter}

\section{Introduction}
\label{sec:intro}

\begin{figure}
\begin{center}
\includegraphics[width=0.3\textwidth]{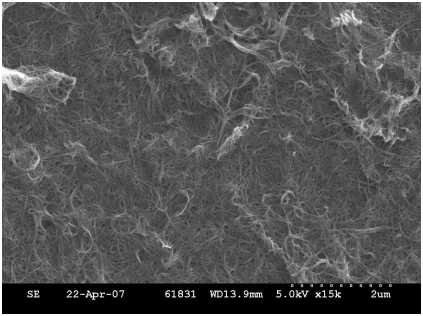} 
\caption{Scanning electron micrograph of a typical carbon nanotube film produced by deposition of a solution containing dispersed nanotube bundles. Image is courtesy of J. N. Coleman's research group, Trinity College Dublin.}
\label{fig:film}
\end{center}
\end{figure}

Carbon nanotubes present remarkable physical properties and since their discovery, have shown much promise in terms of applications \cite{Baughman2002}. 
Nanotube network films are being heavily investigated as the basic building material for a new generation of flexible electronic devices. In particular, thin metallic films are currently the most promising material to enable the construction of large scale flexible electronic displays \cite{Snow2003, Wu2004, Cao2009}. Carbon nanotube (CNT) films are produced by deposition of solutions containing dispersed CNT bundles. Figure \ref{fig:film} shows a typical nanotube network film produced by this technique. In general, it is possible to control the density of the films by controlling the amount of solution deposited.

The fabrication of novel electronic displays requires flexible, transparent, highly conductive films and nanotube networks seem to possess all the right ingredients. Flexibility is a basic property of CNTs. Transparency can be achieved by controlling the density of nanotubes in the network. Finally, while individual CNTs are known to have remarkable transport properties, the electronic transport in network films takes place through individual nanotubes interconnected by junctions. In this case, electrons experience low resistance when travelling along a nanotube, but are faced with significant resistance at the junctions. Furthermore, transparency and conductivity are intimately connected in thin metallic films. A very dense film might show high electrical conductivity but low transparency. On the other hand, a sparse network may show high transparency but poor conductivity. It is essential to search for optimum ways of increasing the conductivity but maintaining a good level of optical transparency. To achieve that, it is necessary to obtain a clear understanding of how the electronic transport occurs in such networked films, a task in which computer models are extremely useful.

Modelling complex structures such as carbon nanotube network films is a huge challenge. A successful theoretical model needs to account for the topology and the disordered nature of the arrays formed by individual nanotubes (and bundles). It is also necessary to account for the electronic coupling between individual components throughout the films, since electronic transport takes place through a tunnelling network spanning the whole film. Finally, it is necessary to account for the electronic structure of individual carbon nanotubes. 

The problem presents two clear length scales. One which accounts for the geometrical aspects of nanotube network films, as well as the complex topology of these structures. We shall refer to this as the macroscopic scale. The other length scale considers that carbon nanotubes are made of individual atoms arranged in the usual hexagonal structure. This shall be referred to as the microscopic scale. Once the relevant length scales have been clearly identified, it is possible to develop a multiscale model which accounts for both with the appropriate accuracy.

The geometrical aspects of individual carbon nanotubes can be properly captured by a simple model where these are represented by rods of specified length and diameter. Choosing a random position and orientation for each rod, within a containing volume, a network of interconnected elements is generated, which mimics the topology of the networks in CNT films. Naturally, percolation theory methods can be used to investigate the transport characteristics of these networks. On the microscopic scale, the electronic structure of a carbon nanotube can be satisfactorily described by simple model Hamiltonians. The tight binding approximation, considering only one orbital per carbon atom, is enough to provide a description of the relevant electronic states. Furthermore, standard Green function methods can be used to describe the electronic structure and electronic transport characteristics across the nanotube networks. 

From a theoretical viewpoint, the description of the transport properties of such networks requires a multi-disciplinary effort involving classical electromagnetism, quantum physics and statistics. In this work we attempt to use tools of all these multidisciplinary fields of study to present a multiscale approach that can describe the electronic transport on carbon nanotube films. It begins with a description of the model utilised to account for the complex topological network formed by CNTs. It will be shown how the connectivity of the network depends on the density of the network as well as the dimensions of individual nanotubes. Considering a limit where the network resistance is exclusively due to nanotube junctions, it will be shown how the proposed model compares with standard percolation theory predictions. Moving on to the microscopic description of the problem, it will be shown how Green function methods can be used to calculate electronic transport on disordered nanotube networks. Finally, by combining both length scales in a single formalism, we show how the resulting multiscale model can be applied to calculate the maximum conductance of network films.

\section{Random Networks of Finite Rods}
\label{sec:geomodel}

The complex structure presented by carbon nanotube network films can be modelled by standard  geometric percolation models. Single nanotubes and small diameter bundles are modelled by  rigid rods of specific length and diameter. The structure of the film itself can be represented by a 3D box which encloses the disordered array of interconnected rods. 
It is important to define a box with appropriate dimensions, such that most rods are located inside the box. It is also important to reproduce other aspects present on the real experiments, such as the placement of metallic electrodes on the sides of the films for electrical resistance measurements.

Once the aspect ratio of the rods, defined as length-to-diameter ratio, and the dimensions of the enclosing box are defined, it is also necessary to specify the total number of rods inside the box, i.e. the number of nanotubes in the film. The total number of rods considered along with the dimensions of the box determine the density of the film. The density of a specific CNT film can also be expressed in terms of its volumetric fraction occupied by nanotubes. The volume fraction  of a typical laboratory produced nanotube film is related to the volume of CNT solution used for deposition \cite{Hu2004, Hu2010}, and can be directly measured. In a computer simulation, the volume fraction depends on the total number of rods distributed inside the box, and is calculated directly for each configuration generated.

\begin{figure}
\begin{center}
\includegraphics[width=0.3\textwidth]{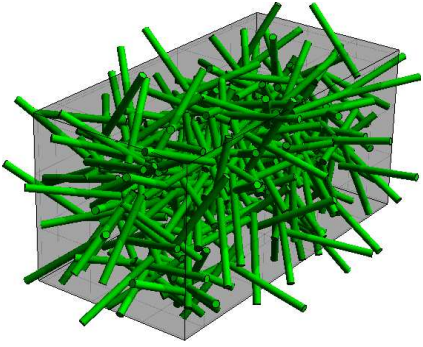} 
\caption{Illustrative representation of rigid rods randomly placed inside a rectangular box. There is a total of $N=300$ identical rods inside the box, each with aspect ratio $\ell/D=20$. }
\label{fig:3d-box}
\end{center}
\end{figure}

Each rod is randomly placed and randomly oriented inside the box. The arbitrary orientation direction is chosen in a 3D space with no constraints, and individual rods are allowed to overlap with other rods. An arbitrary rod might also have part of its volume located outside the box, and this portion is not included when calculating the volume fraction of the film. Each ensemble of randomly distributed/oriented rigid rods gives rise to a complex random network of interconnected 1-dimensional elements, capable of capturing the most important morphological features found on a nanotube network. Figure \ref{fig:3d-box} shows a typical configuration generated with $N=300$ identical rigid rods of length-to-diameter ratio equal to $20$.

It is possible to identify interconnections between rods throughout the network by considering the centre-to-centre distance between each pair of rods. If the smallest centre-to-centre distance between a given pair is smaller than the sum of their radii, then this pair is said to be connected. By calculating the distance between each possible pair of rods in the network, all the connections are mapped and a list of neighbours for each rod is constructed. In general, each rod has an arbitrary number of neighbours, but the number of connections (neighbours) per rod follows some probability distribution with a well defined mean. The average number of connections per node is defined as the connectivity of the network. Networks with a low density of rods, i.e. a small volume fraction, are expected to have low connectivities, while networks with a large volume fraction will have high connectivities. In other words, the density of junctions in a network film depends directly on its volume fraction.

In order to investigate electronic transport across random nanotube networks, it is necessary to include the placement of metallic electrodes in model films. Contacts to electrodes are crucial to electronic transport studies, and must be addressed with caution. In experimental realisations, metallic electrodes are placed at opposing ends of a film and the resistance is measured between these electrodes in a usual two probe method. In this work electrodes are modelled by two opposing faces of the 3D box. A rod crossing one of the specified electrode faces is considered to be in contact with that particular electrode. In this manner, a map of all the rods connected to the electrodes is constructed.

\subsection{Simulation Details}

Individual carbon nanotubes are represented by rigid rods of length $\ell$ and diameter $D$, with a well defined aspect ratio given by $\ell/D$. The film is mimicked by a rectangular cuboid, whose dimensions are defined in terms of the rod length $\ell$. In all the results reported here the box has dimensions given by $2 \ell \times 2 \ell \times 4 \ell$. For each rod, a random position for one of its ends is generated inside the box, and the position of the other end is determined by the orientation of the rod. The total number of rods considered in each film is $N$, and along with $\ell$ and $D$ these are the parameters of the simulations.

The orientation of the rods is chosen at random from all the possible directions in a 3D space. A particular rod can have one of its endings located outside the box, and the portion of the rod located outside of the film is not considered when calculating the volume fraction. Rods are allowed to overlap with other rods, which might lead to an overestimation of the volume fraction, especially in the case of very dense networks. However, this possible overestimation of the volume fraction does not affect the morphology of the  resulting networks. 

Electrodes are located on the  $2 \ell \times 2 \ell$ faces of the box, and rods crossing one of those faces are connected to that electrode. The box is defined with a separation large enough to avoid an individual rod being connected to both electrodes, in order to avoid short circuits.
For a fixed number of rods, and a box of fixed dimensions, a very large number of configurations can be generated by randomly distributing the rods. Therefore, a large number of independent configurations must be considered, and a careful statistical analysis of the results must be carried out. All the results reported in the next section are obtained by averaging a large number of independent realisations, without discarding any configuration.

In order to determine which pairs of rods are interconnected it is necessary to calculate the minimum distance between each rod and all the rods around it. Identifying which rods are in contact with the electrodes also requires testing the criteria for all rods close to the respective faces. Both procedures can be very time consuming from the computational point of view, and the computation time grows very rapidly with the total number of rods considered. In all simulations we have considered networks with a maximum of $1800$ individual rods.

\subsection{Results}

\begin{figure}
\begin{center}
\includegraphics[width=0.4\textwidth]{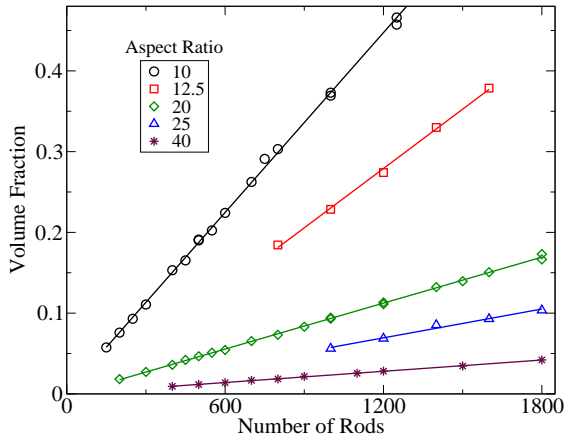} 
\caption{Dependence of the volume fraction on the total number of rods. The lines shown in the graph are linear least-square fits to each data set, and error bars are smaller than symbol sizes.}
\label{fig:Vf-vs-N}
\end{center}
\end{figure}

Let us begin by analysing the relation between the total number of rods and the volume fraction of the films simulated. As previously mentioned, experimental studies have found a direct relationship between the volume of solution deposited and the volume fraction of the resulting network \cite{Hu2004}. 
Figure \ref{fig:Vf-vs-N} shows the relationship between the volume fraction and the total number of rods randomly distributed in the film. The data is shown for different aspect ratios, which is controlled by varying the diameter of the rods. Each data point is an average over several independent configurations and error bars are smaller than symbol sizes. Solid lines shown in the graph are linear least-square fits to each data set, and confirm a direct proportionality between the volume fraction and the total number of rods in the network. 
The existence of such a direct relation between the number of rods and the volume fraction  provides a direct way of comparing and combining simulation results with experimental measurements. On a real CNT film there are several thousand individual nanotubes, whereas computational resources limit simulations to only a few thousand rods. However, morphological properties such as connectivity and density of junctions are directly related to the volume fraction, which connects experiment and simulation at least on a qualitative level. 

\begin{figure}
\begin{center}
\includegraphics[width=0.45\textwidth]{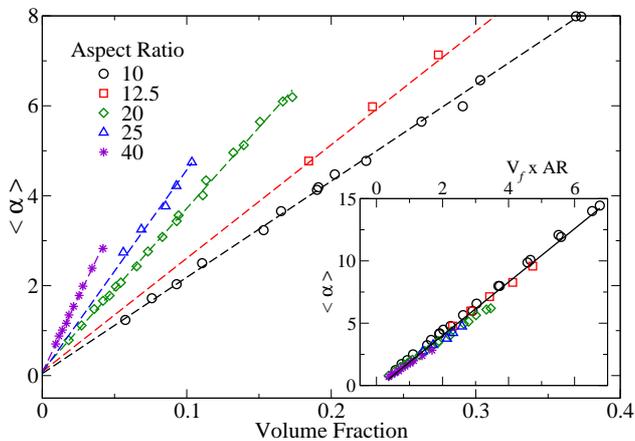}
\caption{Average network connectivity as a function of volume fraction. Inset: Universal scaling of the connectivity with $V_f \times \ell/D$. Error bars are smaller than symbol sizes.}
\label{fig:alpha-vf}
\end{center}
\end{figure}

Now we shall proceed to analyse how the connectivity of the network depends on the concentration of rods. For a film of fixed dimensions and containing rods of identical aspect ratio, it is natural to expect that the average connectivity will increase with the volume fraction. If there are more rods inside a fixed finite volume, more rods will be in contact. On the main panel of figure \ref{fig:alpha-vf} it is shown how the average connectivity per rod $\avga$, increases with the volume fraction $V_f$, for each aspect ratio considered.  From the data it is also clear that for a fixed volume fraction, networks composed of high aspect ratio rods have a higher connectivity when compared to lower aspect ratio rods. This is an interesting feature, which is in accordance with experimental observations that films made of high aspect ratio bundles present a lower overall resistivity \cite{Hu2010}.

Based on a rod percolation model similar to the one considered here, a purely geometric argument has been used to propose that the mean number of junctions per nanotube (or bundle) is expected to behave as \cite{Blighe2008}
\begin{equation}
\label{eq:alpha}
\langle \alpha \rangle = \beta  \times V_f \frac{\langle \ell \rangle}{\langle D \rangle},
\end{equation}
where $\beta$ is a constant independent of the nanotube (or bundle) diameter and length. Also, $V_f$ is the volume fraction, $\langle \ell \rangle$ is the average bundle length, and $\langle D \rangle$ is the average bundle diameter. 
In order to verify if this hypothesis is satisfied by the present model, the average connectivity is plotted as a function of the scaled variable $V_f \times \ell/D$, in the inset of figure \ref{fig:alpha-vf}. Remarkably, the results are in good agreement with the predicted relation, for all the aspect ratios considered. Besides providing a direct verification of equation (\ref{eq:alpha}), from the linear least-square fit on the inset of figure \ref{fig:alpha-vf} the value of the constant $\beta = 2.14 \pm 0.03$ is calculated, which is independent of the other factors in the equation. 

\begin{figure}
\begin{center}
\includegraphics[width=0.4\textwidth]{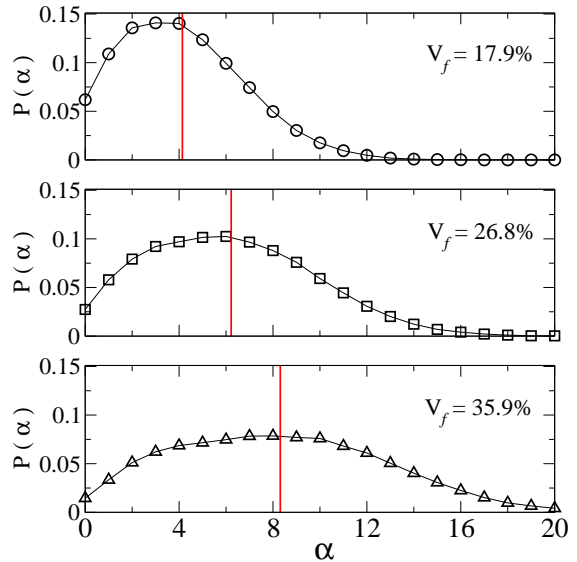} 
\caption{Connectivity probability distribution for random networks of finite rods with aspect ratio $\ell/D=10$, and indicated volume fraction. Vertical solid lines indicate the calculated average connectivity. Each data point is an average over $100$ independent configurations.}
\label{fig:pk-vs-k}
\end{center}
\end{figure}

It is also instructive to analyse the probability distribution of connections per rod in the network. Figure \ref{fig:pk-vs-k} presents the distribution for the probability that an arbitrary rod will have $\alpha$ connections. The data is obtained by considering random configurations with rods of aspect ratio $\ell/D=10$. The vertical solid lines indicate the calculated average connectivity $\avga$ for each volume fraction specified. From the data it can be clearly seen that the average number of connections per rod increases as the concentration of the network is increased, while the probability distribution becomes broader. The broadening of the distribution indicates that as the density of the networks increases, they actually become less homogeneous.

\begin{figure}
\begin{center}
\includegraphics[width=0.45\textwidth]{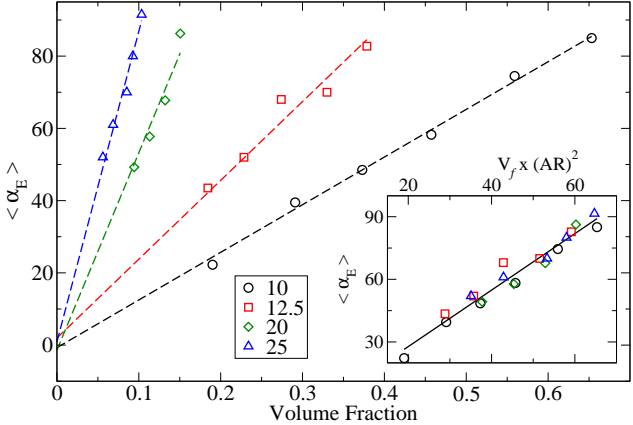} 
\caption{Number of rods connected to electrodes as a function of the volume fraction. Inset: Universal scaling of the electrode connectivity with $V_f \times (\ell/D)^2$.}
\label{fig:alphaE-vf}
\end{center}
\end{figure}

Another important factor extracted from the computer model considered is the number of rods connected to the electrodes. In figure \ref{fig:alphaE-vf} (main panel) it is shown that the number of rods crossing the electrode faces increases with the volume fraction of the film. Similar to the average connectivity of the network, the actual increase of $\avgae$ with $V_f$ depends on the aspect ratio of the rods, albeit in this case the dependence is more pronounced. In order to be useful when compared to experimental data, it is desirable to find a universal scaling dependence for $\avgae$ in terms of the parameters $V_f$, $\ell$ and $D$, similar to the one in equation (\ref{eq:alpha}). Following an empirical approach, it has been found that the relation
\begin{equation}
\langle \alpha_E \rangle = \beta_E \times V_f \left( \frac{\ell}{D} \right)^2,
\label{eq:alphaE}
\end{equation}
provides an excellent collapse of the data on a single straight line, as shown on the inset of figure \ref{fig:alphaE-vf}. Moreover, the proportionality constant is found to be $\beta_E = 1.35 \pm 0.07$. 

In practice, equations (\ref{eq:alpha}) and (\ref{eq:alphaE}) provide numerical values for the connectivity and for the number of connections to the electrodes in a model network film, for any set of parameters chosen. Furthermore, the expressions obtained for $\avga$ and $\avgae$ can be used in combination with experimental data in order to estimate physical properties of real carbon nanotube films \cite{Lyons2008}.
Finally, equations (\ref{eq:alpha}) and (\ref{eq:alphaE}) can also be coupled with electronic transport models to provide calculations for the electrical conductivity of network films. 

In the next section we present a methodology to calculate the resistance of a network composed of Ohmic resistors. This methodology is later coupled with the results presented above, to provide calculations for the resistance of nanotube films from a purely classical point of view. Later on, the results presented above will be coupled with a quantum transport model, providing a more realistic microscopic picture of electronic transport across carbon nanotube films.

\section{Resistive Networks}
\label{sec:resnet}

In carbon nanotube films, the resistance of the nanotubes itself is negligible when compared to the junction resistance. In this case, a nanotube network film resembles a networks of perfect conductors connected by Ohmic resistors.
Calculating the resistance between arbitrary nodes on a lattice of resistors is a problem that has been investigated before. Consider for example a resistive lattice where each vertex contains a resistance $R_0$, i.e. between every two adjacent nodes there is an Ohmic resistor $R_0$. Different methodologies have been developed in order to tackle this problem. 

Venezian \cite{venezian1994} introduced a method to calculate the  resistance between two adjacent nodes on an infinite square grid of equal resistors based on a superposition of potentials and currents of two one-terminal configurations. Atkinson and van Steenwijk  \cite{atkinson1999} generalised the original method to 3D and higher-dimensions. Cserti and collaborators developed a methodology based on the application of lattice Green functions \cite{cserti2000, cserti2002}, which provides values for the equivalent resistance in infinite networks. Building on the method developed by Cserti, it is possible to propose yet another approach specifically aimed at dealing with finite networks of resistors. The methodology consists on the direct application of Ohm's and Kirchoff's circuit laws to finite resistive networks.
When a current is injected in an arbitrary node and extracted from another arbitrary site, the expression for the equivalent resistance is given in terms of the potential difference between these sites. 

Consider a resistive network where between every two adjacent sites there is a resistor $R_0$. The current that can be extracted from a site denoted by $A$ is given by the net current from its nearest neighbours, which can be written as
\begin{equation}
\label{eq:kirchoff}
I_{A} = \frac{1}{R_0} \left[ (V_{A}-V_{A+1}) + (V_{A}-V_{A-1}) + \cdots \right],
\end{equation}
where $V_{A}$ is the electrical potential on site $A$, and $A\pm1$ represent two of the nearest neighbours of site $A$. It is possible to rewrite equation (\ref{eq:kirchoff}) as
\begin{equation}
I_{A} = \frac{1}{R_0} \left[ \alpha_A V_{A} - \sum_{\delta}^{\alpha_A} V_{A+\delta}  \right],
\end{equation}
where $\delta$ represents the nearest neighbours of site $A$, and $\alpha_A$ its total number of neighbours. 
In matrix notation, it can be written in a compact form as 
\begin{equation}
\mathbf I = \frac{1}{R_0} \mathbf M \mathbf V.
\label{eq:kirchoff-matrix}
\end{equation}
Only two elements of $\mathbf I$ are non-zero, the ones corresponding to the injection and extraction of current, being one positive and one negative. 

The matrix denoted by $\mathbf M$ has the general form
\begin{equation}
\mathbf M = \left(
\begin{array}{ccccc}
\alpha_1 & -1 & 0 & 0 & \cdots  \\
-1 & \alpha_2 & -1 & 0 & \cdots \\
0 & -1 & \alpha_3 & -1 & \cdots \\
\vdots & \vdots & \vdots & \vdots &\ddots \\
\end{array}
\right),
\end{equation}
where $\alpha_A$ is again the number of nearest neighbours of site $A$, and $M_{A,B}=-1$ if
$A$ and $B$ are neighbours or zero otherwise. Notice that for a true random lattice, each site has an arbitrary number of neighbours. 

At this point, the matrix equation (\ref{eq:kirchoff-matrix}) is re-arranged to read 
\begin{equation}
\mathbf V = R_0 \mathbf M^{-1} \mathbf I.
\label{eq:kirchoff-matrix-inv}
\end{equation}
By performing the calculation above, we obtain the value of the electrostatic potential in each and every node of the network. Moreover, from $\mathbf V$ one calculates the potential difference across every resistor in the network, which provides a direct way of calculating the equivalent resistance between any two arbitrary points of the network. Furthermore, by constructing the appropriate matrix $\mathbf M$, it is possible to calculate the resistance between arbitrary nodes of any resistive structure, all that is required is to describe the matrix  $\mathbf M$ according to the specific network connectivity.

The resistance between two arbitrary sites $A$ and $B$ is given by
\begin{equation}
R_{A,B} = \frac{V_A - V_B}{I}.
\end{equation}
Which can finally be expressed in terms of specific elements of the inverse of $\mathbf M$ as
\begin{eqnarray} \nonumber
R_{A,B} &=&R_0 ( \; [M^{-1}]_{A,A} + [M^{-1}]_{B,B} \\ &-& [M^{-1}]_{A,B} - [M^{-1}]_{B,A} \; ).
\label{eq:R}
\end{eqnarray}
Therefore, the resistance between two arbitrary points in any type of
lattice can be calculated as long as we know the number of connections, and which sites are interconnected. It is not hard to generalise this result for the case where the resistance between nearest neighbours is not identical, or even for a distribution of different resistances. In order to do so, the $\frac{1}{R_0}$ term in Eq. (\ref{eq:kirchoff}) would have to be replaced by the specific resistance connecting the corresponding sites, and could be absorbed into the definition of $\mathbf M$.

\subsection{Application}
\label{sec:res-comp-net}

The method described above can be directly applied to model carbon nanotube films where the resistance between the two electrodes is calculated through equation (\ref{eq:R}). Generating a random configuration of rods and calculating the connectivities of the arrangement, the matrix $\mathbf{M}$ is written and the resistance between the electrodes is calculated. It is important to recall this approach is valid when the resistance along the nanotubes is very small, and the resistance of the film is dominated by the junctions. 

\begin{figure}
\begin{center}
\includegraphics[width=0.3\textwidth]{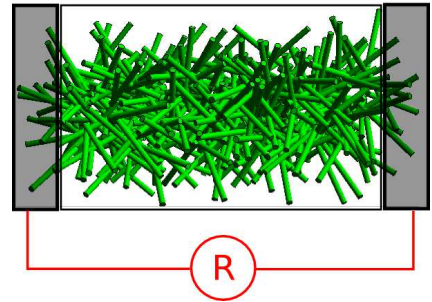} 
\caption{Schematic representation of the model used to calculate the resistance of carbon nanotube networks. Individual CNTs and bundles are represented by rods of finite length. Ohmic resistors $R_0$ are place at each junction. Total resistance  across the film is calculated as the equivalent resistance between electrodes (shaded lateral rectangles). }
\label{fig:film-resistance}
\end{center}
\end{figure}

A random array of nanotubes (represented by finite-length rigid rods) is generated inside a containing box of specified dimensions, and the volume fraction is calculated. The junctions between rods in close proximity are mapped and a list of neighbours of each rod is constructed. Similarly, rods that cross the electrode faces are identified, and that specific electrode is added to their list of neighbours. The inter-tube junction resistances experienced by conduction electrons, are represented by Ohmic resistors ($R_0$), placed between every two rods in contact.  Finally, the total resistance between the electrodes is calculated, and for the purpose of analysis, it is converted into conductance by taking its inverse. Figure \ref{fig:film-resistance} illustrates in a schematic way the set up used in our calculations.

As previously mentioned, nanotube films are commonly produced by deposition of solutions containing dispersed CNTs. The disordered network has a density (and volume fraction) which is directly proportional to the volume of solution used. A small volume of deposited solution will generate networks with a low density of connections and few percolating paths for the charge to move from one electrode to the other. In this case, films with high resistance will be produced. On the other hand, if the volume of solution used is large, it is expected that the films produced will have a low overall resistance, since there will be several percolating paths for the conduction electrons to move across the network. 

It is more convenient to express the charge transport results in terms of the inverse resistance, i.e. conductance, defined simply as $\Gamma = 1/R$. Percolation theory predicts that for small values of $V_f$, the conductance behaves  approximately as 
\begin{equation}
\Gamma \sim (V_f - V_c)^t,
\label{eq:percol}
\end{equation}
where $V_c$ is the critical value of the volume fraction. The critical exponent $t$, depends only on the dimensionality of the embedding space. In our model the rods are randomly oriented in a 3D space, and for this case, the theory predicts $t=1.94$ \cite{Stauffer1992}.  Beyond the percolation threshold, i.e. for networks with a volume fraction considerably larger than the critical value $V_c$, the conductance is found to increase linearly with $V_f$ \cite{Pike1974, Seager1974}.

\begin{figure}
\begin{center}
\includegraphics[width=0.4\textwidth]{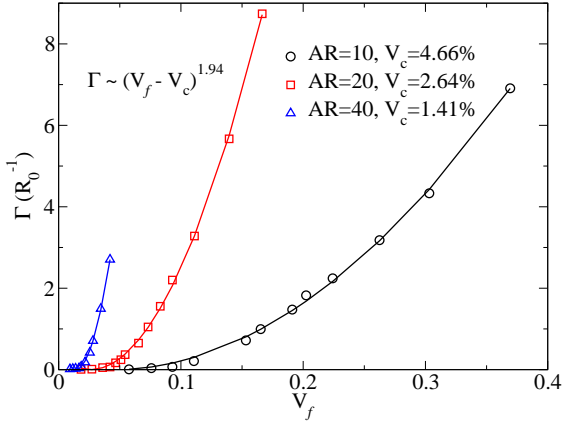} 
\caption{Network conductance as a function of the volume fraction near the percolation region. Solid lines are a fit of equation (\ref{eq:percol}) with the appropriate $3D$ exponent $t=1.94$.}
\label{fig:G-vs-Vf-comp2}
\end{center}
\end{figure}

The behaviour of $\Gamma$ in the percolation region can be analysed in detail in order to verify the agreement with the standard theory. 
In figure \ref{fig:G-vs-Vf-comp2}, we fit equation (\ref{eq:percol}) with the expected scaling exponent $t=1.94$, to the data near the percolation region. The agreement with theory is remarkable, and from the fit we also extract the value of the critical volume fraction for each aspect ratio considered. The calculated values of $V_c$ are inversely proportional to $\ell/D$, which is an intuitive result and is in agreement with previous theoretical predictions \cite{Pike1974, Munson-McGee1991, Celzard1996}.

In section \ref{sec:geomodel} it has been shown how the connectivity of a random network of rods scales linearly with the product of the volume fraction with the aspect ratio of individual rods. It is natural to expect the conductivity of the network to increase with its connectivity, which is indirectly demonstrated by the dependence of the conductance with the volume fraction (see figure \ref{fig:G-vs-Vf-comp2}).
Moreover, it is also possible to analyse the dependence of the film conductance with the density of junctions, which is defined as the total number of junctions per unit volume, and can be written in terms of the network connectivity as 
\begin{equation}
N_J = \frac{\langle \alpha \rangle N}{2} \frac{1}{V},
\end{equation}
where $N$ is the total number of rods inside the box of volume $V$. 

\begin{figure}
\begin{center}
\includegraphics[width=0.4\textwidth]{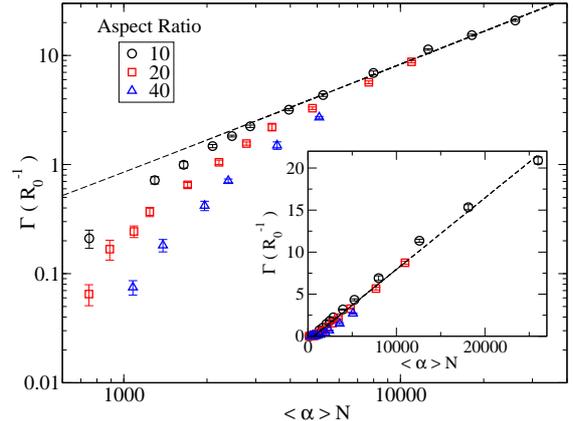} 
\caption{Film conductance increases linearly with density of connections for large enough densities. Log-log plot shows convergence to linear regime.}
\label{fig:G-vs-Nj}
\end{center}
\end{figure}

Figure \ref{fig:G-vs-Nj} shows a plot of the network conductance as a function of $\langle \alpha \rangle N = 2 N_J V$ on a log-log scale, and on a linear scale on the inset, for three different aspect ratios. 
Once again our calculations are limited by the computational resources required for simulations with the rods of large aspect ratio. However, it is possible to infer from the log-log plot that, at least for high enough concentrations of junctions, the conductance seem to converge to the linear dependence represented by the dashed straight line. Furthermore, in the inset we show the excellent collapse of all the data to the same straight line in a linear scale $\Gamma/R_0^{-1} = m \times 2 N_J V$.
In combination with experimental measurements, the model described above has been used to estimate the inter-tube junction resistance \cite{Lyons2008}. The result obtained is in broad agreement with independent experimental results \cite{Fuhrer2000, Nirmalraj2009}. Nonetheless, it is important to point out that in real CNT films the junction resistance can vary widely, depending on the chirality of the nanotubes involved as well as the diameter of the bundles, and an exact determination of these contact resistances is still an open problem.



\section{Microscopic Modelling}
\label{sec:model}

The approach presented in the previous sections analyses the system in a macroscopic scale. Individual CNTs were modelled as perfectly conducting rigid rods of finite aspect ratio, and tunnel barriers at nanotube junctions were mimicked by Ohmic resistors. The film resistance was calculated as the equivalent resistance between opposing electrodes separated by a  disordered network of identical resistors. As previously mentioned, the method provides results in reasonably good agreement with experiments and previous percolation model studies.

It is also possible to approach electronic transport problems from an atomistic viewpoint, most notably by means of {\it ab-initio} density functional theory calculations (DFT) coupled with non-equilibrium Green function methods \cite{Brandbyge2002,Rocha2005}. 
However, the computational cost of these calculations can be considerably high. Moreover, all current DFT implementations require at least some degree of periodicity in order to handle large systems ($\approx 10^3$ atoms), which unfortunately is clearly not present in disordered CNT networks.
Nonetheless, {\it ab-initio} methods have been used to investigate electrical resistance at single wall nanotube (SWNT) junctions \cite{Yoon2001}, and more recent studies have extended these results considering the  presence of O$_2$ and N$_2$ molecules \cite{Mowbray2009} near the junction. 
The use of first principles methods for obtaining the quantum conductance between individual nanotubes is able to shed some light on the quality of the junction, but it is unable to reproduce the general observed features in network films because it is not capable of including disorder effects to the scale required.
The alternative of carrying out a fully atomistic transport calculation within a heavily disordered environment is very desirable but it is currently too computationally demanding. It is therefore necessary to compromise if one wishes to combine the two features. 

One possible approach consists of using computationally inexpensive semi-empirical methods to describe the electronic structure of carbon nanotubes. Tight binding model Hamiltonians are a very efficient and convenient way to describe electronic structure of materials in general, and are well suited to deal with carbon nanotubes. By describing the electronic structure of individual nanotubes with a simplistic model Hamiltonian, we can afford to account for disorder effects by considering hundreds, if not thousands, of CNTs, and  several possible configurations for the structure of the random tunnelling networks formed in the films. Furthermore, this approach is particularly well suited to treat electronic transport across networks on a purely ballistic regime. By neglecting all sources of decoherence-inducing scattering, it is possible to calculate the conductance of ideal network films. The conductance calculated in such idealised scenario represents an upper bound for the transport performance of real films, and can be used as an indication of how much room there is for improving the conductivity of these materials.  

\subsection{Tight Binding Hamiltonian and Green Functions}

A general semi-empirical tight binding Hamiltonian can be written as
\begin{equation}
{\mathcal H_\nu} = \sum_{i} |\nu, i \rangle \epsilon_0 \langle \nu, i | + \sum_{ \langle i,j \rangle } |\nu, i \rangle \gamma_0 \langle \nu, j |,
\label{TB-H}
\end{equation}
where $|\nu, i \rangle$ represents an electron on atom $i$ of a wire labelled by $\nu$, $\epsilon_0$ is the on-site energy, $\gamma_0$ is the hopping integral parameter, and the sum $\langle i,j \rangle$ is over nearest-neighbouring atoms. This general form can describe several types of nanostructures, depending on the choice of atomic structure and on the orbital degrees of freedom represented by the states $| i \rangle$. Associated with the Hamiltonian ${\mathcal H}_\nu$, there are the retarded($+$) and advanced($-$) single-particle Green functions, which are defined as
\begin{equation}
{\mathcal G}^{\pm} = [E \pm \imath \eta - {\mathcal H_\nu}]^{-1},
\label{eq:GF}
\end{equation}
where $\eta$ is a small positive imaginary part added to the electronic energy $E$ in order to avoid singularities at eigenvalues of the Hamiltonian.

Electronic transport calculations typically require the presence of particle reservoirs, which represent source and drain for the conduction electrons. In the case of transport simulations of carbon nanotube networks, semi-infinite nanotubes are employed to mimic the presence of particle reservoirs. These semi-infinite CNTs are solely described by the GF elements relative to   the atoms on its final unit cell. In order to calculate surface GF elements of semi-infinite nanotubes, we apply efficient recursion methods, which are numerically stable and computationally inexpensive.

\subsubsection{Network Green Functions}

Following the general method applied on electronic transport calculations, the system is separated into three parts: two leads and a central region \cite{Caroli1971, Datta1997}. The major difference between our particular approach and the usual one is that each one of the three parts of the system are themselves composed of several wires. The leads are represented by a number of semi-infinite nanotubes, which act as  particle reservoirs. Meanwhile, the central region of the system consists of the disordered network itself, represented by an ensemble of finite-sized CNTs.
We begin by constructing the Green function matrices of a completely disconnected system. The inter-tube junctions are introduced in the form of electronic hopping terms between CNTs in the network, and between the network and the electrodes. Finally, the GF of a fully connected network of nanotubes is calculated with the Dyson equation.

Beginning from a disconnected tripartite system, one can write the complete Green function as a super matrix of the form
\begin{equation}
g^\pm = \left(
\begin{array}{ccc}
g^\pm_L & 0 & 0  \\
0 & g^\pm_C & 0  \\
0 & 0 & g^\pm_R  \\
\end{array}
\right) ,
\end{equation}
where $g^\pm_{L,R}$ are the GF matrix of the left and right electrodes, and $g^\pm_C$ is the GF of the central scattering region. At this stage, the network is an ensemble of non-interacting nanotubes. 

In this initial description of the disconnected system, each of the matrices in $g^\pm$ are in block diagonal form, with one block for each individual CNT. The electrodes act as charge reservoirs which are represented by semi-infinite nanotubes. In other words, whenever a CNT is contacted to either electrode we will represent it by a semi-infinite object. Therefore,
there are $\avgae$ diagonal blocks given by surface GF of semi-infinite carbon nanotubes. In general, the electrode GFs can be written as
\begin{equation}
g^\pm_{L,R}(E) = \left(
\begin{array}{cccc}
\mathcal{S}^\pm(E) & 0 & 0  & \cdots \\
0 & \mathcal{S}^\pm(E) & 0  & \cdots \\
0 & 0 & \mathcal{S}^\pm(E)  & \cdots \\
\vdots & \vdots & \vdots & \ddots \\
\end{array}
\right) ,
\label{gLR}
\end{equation}
where $ \mathcal{S}^\pm(E)$ are advanced (or retarded) Green functions of the surface elements   of semi-infinite wires, which mimic the particle reservoirs.

Similarly, the GF of the central scattering region consisting of a disconnected network with $N$ wnanotubes is represented by a block diagonal matrix with $N$ blocks, and it is written as 
\begin{equation}
g^\pm_{C}(E) = \left(
\begin{array}{cccc}
\mathcal{G}^\pm(E) & 0 & 0 & \cdots \\
0 & \mathcal{G}^\pm(E) & 0 & \cdots \\
0 & 0 & \mathcal{G}^\pm(E) & \cdots \\
\vdots & \vdots & \vdots & \ddots \\
\end{array}
\right) ,
\label{gC}
\end{equation}
where we recall that $\mathcal{G}^\pm(E)$ is the Green function of a finite-sized pristine monatomic linear chain or carbon nanotube. In general, each diagonal block $\mathcal{G}^\pm(E)$ is a matrix whose dimension equals the total number of atoms on the individual wires. 


Individual nanotubes are connected by introducing a perturbing potential bridging any two CNTs that should be in contact. This inter-tube coupling appears in the form of an electronic hopping term $\gamma^\prime$ in the Hamiltonian of the complete system, and is given by
\begin{equation}
 V_{\nu,\nu^\prime} = | \nu, j \rangle \gamma^\prime \langle \nu^\prime , j^\prime |,
\end{equation}
where $\nu$ and $\nu^\prime$ label the connecting wires, and $j$ and $j^\prime$ identify the atoms in each one. Notice that $\nu$ and $\nu^\prime$ must be different while $j$ and $j^\prime$ are allowed to have the same value.

These individual hopping terms can be grouped together in the form of a general perturbation potential written as
\begin{equation}
V = \left(
\begin{array}{ccc}
0 & V_{LC} & 0  \\
V^\dagger_{LC} & V_{CC} & V_{CR}  \\
0 & V^\dagger_{CR} & 0  \\
\end{array}
\right) ,
\end{equation}
where once again each term represents a matrix itself.

On the construction of the perturbing potential we assume that semi-infinite CNTs in the same electrode do not interact with each other. Furthermore, we assume that the central scattering region is large enough, such that there can be no direct hopping terms between semi-infinite wires located on different electrodes. 
The first assumption can be easily relaxed, and it should not have any significant effects on the results.
The second assumption, however, should not be relaxed. The physical meaning of the second constraint, i.e., $[V_{LR}]=[V_{RL}]=[0]$, is to avoid short circuits on the system, which could manifest as spuriously high values for the conductance, and significantly affect the results.

The term $V_{CC}$ is by far the most significant one, and also the one that relies at the heart of our methodology. In our model, $[V_{CC}]$ is a matrix constructed from a random network, generated from an ensemble of nodes with a given average number of connections per node. Equivalently, a random graph can also be defined by its total number of sites $N$, and the total number of connections on the graph $\avga N /2$. 
The process to construct $V_{CC}$ is as follows:
\begin{enumerate}
\item Select two unconnected wires at random: $\nu, \nu^\prime$ ;
\item Select one random atom in each wire to carry on the interaction: $j, j^\prime$ ;
\item Make  $\langle \nu, j | V_{CC} | \nu^\prime, j^\prime \rangle = \gamma^\prime$ ;
\item Repeat until total number of connections equals  $\avga N /2$. 
\end{enumerate}
A similar procedure is also applied to construct $[V_{LC}]$ and $[V_{CR}]$, where the interaction of the electrodes with the finite-sized nanotubes is chosen by a randomised process. The inter-tube hopping term, represented as $\gamma^\prime$ is defined in analogy with the usual tight binding hopping integral parameter $\gamma_0$, and roughly represents the coupling between individual atoms in interconnected CNTs.

Remember that Equations (\ref{gLR}) and (\ref{gC}) describe the GF of the completely disconnected network. The Green function of the interconnected disordered systems can easily be calculated with Dyson's equation,
\begin{equation}
G^\pm = g^\pm + g^\pm V \left( \hat 1  - g^\pm V \right)^{-1} g^\pm ,
\label{eq:dyson1}
\end{equation}
and can be written in matrix form as
\begin{equation}
G^\pm =  \left(
\begin{array}{ccc}
G^\pm_L & G^\pm_{LC} & 0  \\
G^\pm_{CL} & G^\pm_C & G^\pm_{CR}  \\
0 & G^\pm_{RC} & G^\pm_R  \\
\end{array}
\right).
\label{eq:Gpm}
\end{equation}
Equation \ref{eq:Gpm}, is the Green function of the whole interconnected network, including left- and right-hand side electrodes, central scattering region, and propagators between these three regions. Therefore, $G^\pm$ provides all necessary information with respect to the electronic structure of the disordered network, and can be used to calculate the electronic conductance of the system with the Kubo formula.

\subsection{Kubo Formula Applied to Networks}
\label{sec:conductance}

The zero-bias conductance of the system is calculated with the Kubo formula, which provides a simple expression for the conductance by calculating the net electronic current across a reference plane in the system, often referred to as the cleavage plane, which is located between any two parts of the system. The current is expressed in terms of a few Green function matrix elements of the system, and the calculation becomes simple enough. 

\begin{figure}
\begin{center}
\includegraphics[width=0.4\textwidth]{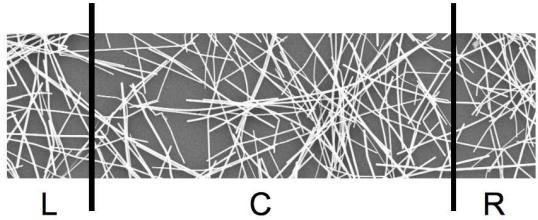}
\caption{Schematic representation of the system, divided in left and right leads (L,R), and a central scattering region (C). The cleavage plane is placed between C and R.}
\label{fig:kubo}
\end{center}
\end{figure}

In analogy to the standard formalism first introduced by Caroli et al. \cite{Caroli1971}, the system is divided in three parts. The location of the cleavage plane is arbitrary and can be chosen in the most convenient location for the case in question. Here it is placed between the central region and the right lead, as illustrated in Fig. \ref{fig:kubo}. Translating Fig. \ref{fig:kubo} into words, regions L and R contain $\langle \alpha_E \rangle$ semi-infinite nanotubes, and region C contains a disordered array of $N$ finite CNTs, forming a random network with an average number of connections per tube given by $\langle \alpha \rangle$. 

Having placed the cleavage plane between the regions C and R, the Kubo formula for the zero-bias conductance is written as \cite{Mathon1997a,Mathon1997b}
\begin{eqnarray}
\label{eq:kubo}
\nonumber
\Gamma(E_F) &=& \left( \frac{4 e^2}{h} \right) \Re e \{ \mathrm{Tr}
[ \tilde G_{C}  V_{CR} \tilde G_{R}  V_{CR}^\dagger \\ &-& V_{CR} \tilde G_{RC} V_{CR} \tilde G_{RC} ] \},
\end{eqnarray}
where all the GFs are calculated at the Fermi energy $E_F$, and $\tilde G(E)$ are defined as
\begin{equation}
\tilde G(E) = \frac{1}{2 \imath} \left[ G^-(E) - G^+(E) \right].
\end{equation}

\subsubsection{Reducing Green Function Matrices}

Even though it is possible to write an explicit expression for the Green function of the interconnected network as in Equation (\ref{eq:Gpm}), from which the matrix elements required by the Kubo formula in Equation (\ref{eq:kubo}) can be obtained, it can still be an extremely challenging task. The difficulty arrises from the size of the matrices involved. Consider for example, a network with $N=100$ wires, where each wire has $L=1000$ atoms. In this case, $g^\pm_C$ is a matrix of $(N \times L)^2 = 10^{10}$ elements, and $[V_{CC}]$ is a matrix of same dimension. Storing matrices of this size on the memory of computer clusters becomes a demanding computational task in itself, and inverting matrices of this dimension takes a considerable time, even if one applies advanced linear algebra routines. 

The solution to this limitation is facilitated by two general properties of the system. The matrix $g^\pm$ is block diagonal and $V$ is a sparse matrix. Therefore, the product $g^\pm V$ on Equation (\ref{eq:dyson1}) is also sparse, or at least has many vanishing terms that will not affect $G^\pm$, i.e., several of the matrix elements in $G^\pm$ do not depend on the perturbation, whenever the corresponding element vanishes.
In practical terms, the corresponding row/column in the matrix $G_C^\pm$ will not involve any $V$-dependent terms and is simply given by ${\mathcal G}^\pm$. 
A connection between atom $j$ of tube $\nu$ and atom $j^\prime$ of tube $\nu^\prime$ will only involve the corresponding matrix elements of the $g$-matrix. Therefore, if a given CNT is connected to three neighbours at intra-tube sites $i$, $j$ and $k$, the matrix elements ${\mathcal G}^\pm_{i,i}$, ${\mathcal G}^\pm_{j,j}$,  ${\mathcal G}^\pm_{k,k}$, ${\mathcal G}^\pm_{i,j,}$, ${\mathcal G}^\pm_{i,k}$, ${\mathcal G}^\pm_{j,k}$ (and the respective adjoint elements) are the only matrix elements appearing in the expression for the network GF of Eq.(\ref{eq:dyson1}). This can be an enormous simplification when compared with the full diagonalisation method. Rather than inverting enormous matrices of dimension of ${N} \times L$, this technique allows us to express the network GF in terms of inverse operations of much smaller matrices, and whose sizes are defined by the total number of connections in the system, i.e. matrices of order $N \times \langle \alpha \rangle$. 

Therefore, because of the particular structure in Dyson's equation, it is possible to reduce the Green function matrices, such that only the elements affected by the perturbation $V$ are stored. Consider, for example, a CNT with several thousand atoms. If only $20$ of these atoms interact with neighbouring nanotubes, then only the elements corresponding to these atoms and the propagators between them need to be stored, since only these will be affected by the perturbation. In this case we can perform calculations up to a thousand wires, each with a few thousand atoms.

\section{Multiscale Model of CNT Networks}

The results from the geometrical modelling presented in section \ref{sec:geomodel} can now be combined with our quantum transport model, providing a multiscale approach to the study of electronic transport in carbon nanotube network films. In particular, by considering an idealised ballistic transport regime it is possible to evaluate the upper limit for the conductance of CNT network films. It has been argued that the construction of transparent, flexible displays requires thin films with an electrical conductivity $\sigma_{dc} \ge 6 \times 10^5$ S/m \cite{Doherty2009}. However, the production of CNT films with such high conductivities has been very difficult to achieve, and it is possible that nanotube networks are not the most suitable material to be used in the next generation of flexible displays \cite{Pereira2009, Segal2009}.

In section \ref{sec:geomodel} results for the connectivity of disordered networks of rods were presented. Universal relations for the dependence of the network connectivity with experimentally accessible quantities such as the volume fraction of the films and the aspect ratio of individual wires, were obtained. 
In particular, it was found that for a random network of rods with individual length $\ell$ and diameter $D$, inside a rectangular cuboid film of dimensions given by $2 \ell \times 2 \ell \times 4 \ell$, the average connectivity per rod is given by
\begin{equation}
\avga = (2.14 \pm 0.03) \times V_f \frac{\ell}{D}.
\label{eq:avga4}
\end{equation}
Furthermore, by considering that metallic electrodes are located on the opposing square faces of the film, and that any rod crossing these surfaces is taken to be in contact to that specific electrode, we found that the number of rods crossing each electrode can be written as
\begin{equation}
\avgae = (1.35 \pm 0.07) \times V_f \left( \frac{\ell}{D} \right)^2.
\label{eq:avgae4}
\end{equation}
Therefore, once the aspect ratio of the wires and the volume fraction of the film are chosen, $\avga$ and $\avgae$ are automatically defined. These parameters are then fed into the general quantum transport model presented in the previous section, which yields the conductance for a network with the specified geometrical properties. Finally, since we have considered disordered networks to be defined inside a film of specified geometry, it is then possible to convert the network conductance into film conductivity. A result which can be directly compared with experimental measurements.

Let us begin by investigating the conductance of networks made of $(5,5)$ carbon nanotubes, which have a metallic band structure. The diameter of a $(5,5)$ CNT is $\approx 0.68$ nm, while the length depends on the number of unit cells considered. In a $(5,5)$ SWNT, each unit cell has $4 \times 5 = 20$ atoms. We consider nanotubes of aspect ratio equal to $11,16.5,22,27.4$, which correspond to lengths of $7.5$ nm, $11.2$ nm, $14.9$ nm, $18.6$ nm, respectively. For each of these aspect ratios, it is possible to vary the total number of carbon nanotubes in the films, such that the network volume fraction ranges from $0.065$ up to $0.325$. Moreover, through equations (\ref{eq:avga4}) and (\ref{eq:avgae4}), the combinations of $\ell/D$ with $V_f$ provide networks with $\avga$ in the range $3.6 - 7.5$, and $\avgae = 27 - 77$.

\begin{figure}
\begin{center}
\includegraphics[width=0.45\textwidth]{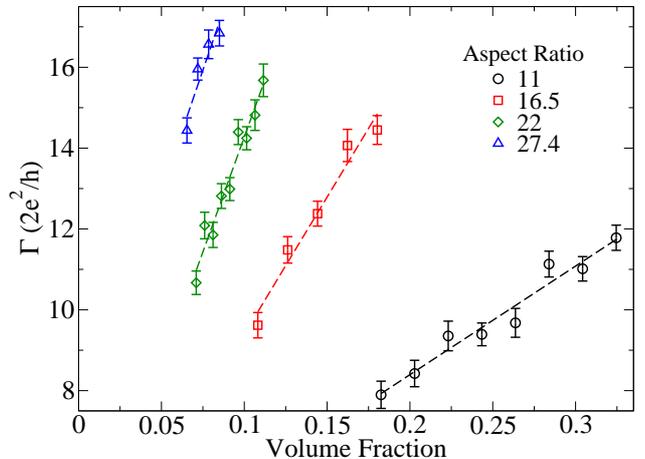}
\caption{Conductance versus volume fraction for networks of $(5,5)$ armchair SWNTs. The legends indicate the aspect ratio of each data set. Higher aspect ratio CNTs provide better conductance, in accordance with experiments \cite{Hecht2006}.}
\label{fig:cond-vs-VF}
\end{center}
\end{figure}

In figure \ref{fig:cond-vs-VF} it is shown how the network conductance depends on the volume fraction for the respective aspect ratios considered. Each data point is an average over at least $50$ independent configurations, and the error bars represent the statistical error in the ensemble. 
Similarly to the resistive network results presented in section \ref{sec:res-comp-net}, there is a clear
improvement in the conductance with a higher concentration of nanotubes. Moreover, it is also clear that higher network conductance values are achieved when the film is composed of nanotubes with larger aspect ratios. The dependence of the film conductance on the aspect ratio of individual nanotubes is in agreement with previous experimental studies \cite{Hecht2006, Shin2009}. 
One possible way to account for this trend is to consider that, on average, a longer nanotube will have a larger distance between its junctions with other tubes. Therefore, on nanotubes with higher $\ell/D$ ratio the electrons will, on average, travel a longer distance along the tubes before experiencing heavy scattering caused by the inter-tube junctions. 

\begin{figure}
\begin{center}
\includegraphics[width=0.45\textwidth]{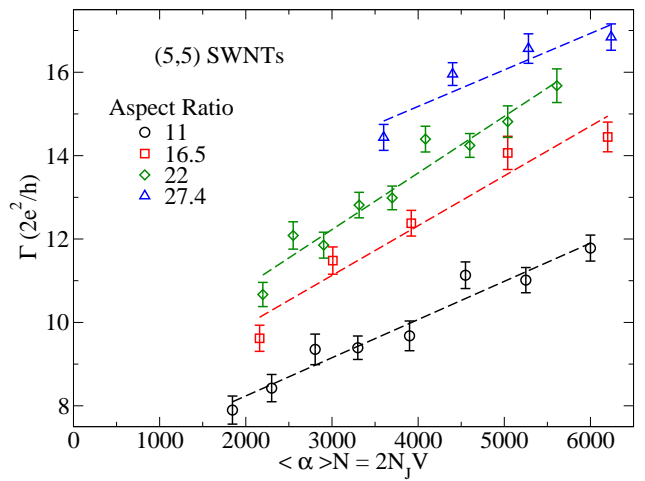}
\caption{Averaged film conductance as a function of total number of nanotube junctions. Dashed straight lines are least-square linear fits to simulation data, in accordance with experimental results of ref. \cite{Lyons2008}. }
\label{fig:cond-alpha-Nj}
\end{center}
\end{figure}

It is also possible to analyse how the conductance depends on the total number of junctions in a network film of volume $V$, which is defined as $N_J V = \avga N/2$. Figure \ref{fig:cond-alpha-Nj} shows the averaged network conductance as a function of the total number of junctions in simulated films. The dashed lines are least-square fits to each data set, and show a linear dependence of the conductance with the number of junctions.
We again see an increase in the overall film conductance as the aspect ratio of individual nanotubes is increased. Moreover, the model reproduces the approximate linear trend observed in experimental reports \cite{Lyons2008}.

It is possible to convert the calculated network conductance into film conductivity by considering the geometry of the films.  Employing the usual relation between conductance and conductivity, it is possible to write 
\begin{equation}
\sigma = \Gamma \frac{L}{A} = \Gamma \frac{4\ell}{4\ell^2} = \frac{\Gamma}{\ell}.
\end{equation}
At this point, it is important to remember that $\Gamma$ represents the conductance of an idealised network of finite-sized carbon nanotubes, in a fully ballistic electronic transport regime. In the calculation of the conductance, all decoherence-inducing factors (certainly present in experimental realisations) were neglected. The calculated conductance is an upper limit for the conductance of a real film made of similar CNTs. The conductivity calculated  $\sigma$, also represents an upper bound for the maximum electrical conductivity of real nanotube network films. 

\begin{figure}
\begin{center}
\includegraphics[width=0.45\textwidth]{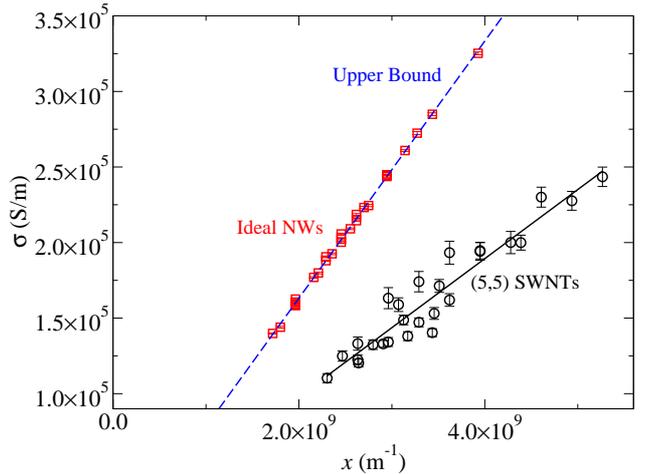}
\caption{Maximum conductivity of carbon nanotube films as a function of scaled variable $x=V_f \times \ell/D^2$.}
\label{fig:sigma_u}
\end{center}
\end{figure}

Based on an empirical analysis of the conductivity data, it is found that it scales with a combined variable of the parameters $V_f$, $\ell$ and $D$. This dependence allows for a direct comparison of experimental data against this superior limit estimate. A comparison between experimental and simulation results, provides a direct indication of how much it is possible to further improve the transport performance of the laboratory produced films.
Defining a scaled variable $x=V_f \times \ell / D^2$, figure \ref{fig:sigma_u} shows the dependence of the film conductivity (in units of siemens per meter) with $x$. The data points represented by circles correspond to nanotubes of varying aspect ratio, and includes all the data in figure \ref{fig:cond-alpha-Nj}. 
The data corresponding to networks of finite-sized $(5,5)$ SWNTs shows considerable dispersion, even after averaging over several configurations. Nonetheless, it is possible to identify a general trend on the data which stresses the linear dependence of the conductivity with $x$, represented by the least-squares linear fit indicated by the dashed (black) line.
The dispersion on the data for CNTs is caused by the finite-length of the nanotubes considered. Since the Green function of independent nanotubes is calculated numerically, it quickly becomes too computationally demanding to increase the length of the nanotubes much further. 

In order to overcome this computational limitation, one can investigate the conductance of networks made of idealised nanowires, represented by linear monatomic chains. In order to compare  the results with carbon nanotubes, an orbital degeneracy is introduced in the chain Hamiltonian such that there are two independent conducting channels. 
These monatomic chains present a maximum conductance of $4 e^2/h$ at the Fermi energy, just  like carbon nanotubes. It is possible to write an analytic expression for the Green function of linear chains \cite{Economou2006}, and very long chains can be simulated with little computational effort. In order to minimise the finite-size effects on the local density of states, chains with $L=16 000$ atoms where considered. 

Also shown in figure \ref{fig:sigma_u} is data obtained for the conductivity of networks composed of ideal nanowires (squares), with varying volume fraction and aspect ratio. In this case we can see a much smaller dispersion on the calculated conductivity. Furthermore, by fitting a straight line to the nanowire conductivity we find
\begin{equation}
\sigma_u = M \times 4.25 \times 10^{-5} \times V_f \frac{\ell}{D^2} \quad \mathrm{( in \; S/m)},
\label{eq:sigmau}
\end{equation}
where the factor $M$ accounts for the number of conduction channels in the wires, for the case considered here $M=2$.
Because of all the idealisations applied to our calculations of the conductivity, the expression above is deemed to be the upper limit for the conductivity of random network films. 

It is instructive to test the expression in equation (\ref{eq:sigmau}) for typical SWNT values, namely $\ell = 1 \mu {\rm m}$ and $d = 1.2 \, {\rm nm}$. For $V_f = 30\%$, the predicted upper bound would be $\sigma_u = 1.8 \times 10^7 {\rm S/m}$ if CNTs of these dimensions could be fully dispersed to form the network. However, carbon nanotubes are known to bundle together, which means that in reality wire diameters are considerably larger. On the other hand, larger-diameter bundles have more nanotubes on the surface leading to more current-carrying channels per wire. Taking all this into consideration, we can compare our expression with the highest-conductivity case reported so far ($V_f=30\%, \ell = 5 \, {\rm \mu m}, d = 20 \, {\rm nm}$) \cite{Wu2004, Geng2007, Doherty2009}. Our prediction of $\sigma_u = 9 \times 10^6 \, {\rm S/m}$ is only one order of magnitude superior to the measured value of $\sigma = 6 \times 10^5 \, {\rm S/m}$. Bearing in mind that the upper bound here obtained assumes a number of ideal conditions that are experimentally unavoidable, this might be a clear indication that we are approaching a saturation point in the conductivity of nanotube network films. 

Finally, although our focus has been on disordered networks comprised of carbon nanotubes, we can extend our results to deal with other wires. This could represent the case of networks made of other conducting materials, such as noble-metal nanowires, for instance. In this case the number of conducting channels M depends linearly on the wire diameter and the overall conductivity of the network is likely to scale inversely with $D$, with a proportionality constant that depends on the specifics of the material in question. If the conductivity of CNT network films is approaching its saturation point, it is likely that wires other than nanotubes may occupy the post of ideal components for disordered network films. Indeed, it has been shown recently \cite{Nirmalraj2009} that silver nanowire network thin films might be a strong contender in the race for the next generation of transparent electrodes. 

\section{Concluding Remarks}

In this work we have shown how a relatively simple methodology can be applied to model the complex features present in the morphological structure of carbon nanotube network films. 
A multiscale approach has been applied to deal with the despairingly different length scales present. From a macroscopic point of view, individual nanotubes are modelled as rigid rods of finite length and diameter. A network of interconnected rods represents the film and the volume fraction of these networks allows for a connection between theory and experiment. On the other hand, from a microscopic approach carbon nanotubes are described by a semi-empirical Hamiltonian, and Green function methods can be applied to calculate the electronic conductance. Merging both scales in a unique multiscale approach is possible when the connectivities derived from the macroscopic model are used as input into the microscopic conductance calculation.

The volumetric fraction of the film occupied by rods was found to be linearly dependent on the number of rigid rods in the network. The average number of connections per rod in a random network was found to scale universally with a combined variable given by the product of the volume fraction with the aspect ratio of the individual rods considered. Similarly, the number of rods connected to electrodes placed in opposing faces of the film was found to scale with the volume fraction times the square of the aspect ratio.
We have presented a method, based on previously developed methodologies, to calculate the equivalent resistance between arbitrary points on a network of resistors. The rigid rods percolation model was combined with resistance calculations in order to obtain the conductance of model carbon nanotube network films. We have shown how our results are in perfect agreement with standard percolation theory, and how the conductance of nanotube films is affected by the morphology of the networks. Direct comparison of simulation results with experimental measurements are found to be in agreement, and provide estimates for the junction resistance in real world nanotube films. These estimates are compatible with independent measurements reported in the literature.

We have shown that an electronic structure description based on semi-empirical tight binding Hamiltonians is capable of accounting for the heavily disordered environments found on nanotube networks. 
In conjunction with results from the geometrical percolation model, we have presented simulation results for the conductance of single-walled carbon nanotube networks.
Finally, by considering a fully ballistic transport regime on films of ideal nanowires, and comparing the ensuing results with typical experimental evaluations we have shown that carbon nanotube network films might be approaching their performance limit, and that perhaps metallic nanowires represent a more promising alternative to future developments.

\bibliographystyle{unsrt} 
\bibliography{library}

\begin{thebibliography}{10}

\bibitem{Baughman2002}
R.H. Baughman, A.A. Zakhidov, and W.A. de~Heer.
\newblock {Carbon nanotubes-the route toward applications.}
\newblock {\em Science}, 297(5582):787--92, August 2002.

\bibitem{Snow2003}
E.S. Snow, J.P. Novak, P.M. Campbell, and D.~Park.
\newblock {Random networks of carbon nanotubes as an electronic material}.
\newblock {\em Applied Physics Letters}, 82(13):2145, 2003.

\bibitem{Wu2004}
Z.~Wu, Z.~Chen, X.~Du, J.M. Logan, J.~Sippel, M.~Nikolou, K.~Kamaras, J.R.
  Reynolds, D.B. Tanner, A.F. Hebard, and A.G. Rinzler.
\newblock {Transparent, conductive carbon nanotube films.}
\newblock {\em Science}, 305(5688):1273, August 2004.

\bibitem{Cao2009}
Q.~Cao and J.A. Rogers.
\newblock {Ultrathin Films of Single-Walled Carbon Nanotubes for Electronics
  and Sensors: A Review of Fundamental and Applied Aspects}.
\newblock {\em Advanced Materials}, 21:29--53, January 2009.

\bibitem{Hu2004}
L.~Hu, D.~S. Hecht, and G.~Gruner.
\newblock {Percolation in Transparent and Conducting Carbon Nanotube Networks}.
\newblock {\em Nano Letters}, 4(12):2513--2517, December 2004.

\bibitem{Hu2010}
L.~Hu, D.~S. Hecht, and G.~Gruner.
\newblock {Carbon Nanotube Thin Films: Fabrication, Properties, and
  Applications}.
\newblock {\em Chemical Reviews}, 110(10):5790--5844, July 2010.

\bibitem{Blighe2008}
F.~Blighe, P.E. Lyons, S.~De, W.J. Blau, and J.N. Coleman.
\newblock {On the factors controlling the mechanical properties of nanotube
  films}.
\newblock {\em Carbon}, 46:41--47, January 2008.

\bibitem{Lyons2008}
P.E. Lyons, S.~De, F.~Blighe, V.~Nicolosi, L.F.C. Pereira, M.S. Ferreira, and
  J.N. Coleman.
\newblock {The relationship between network morphology and conductivity in
  nanotube films}.
\newblock {\em Journal of Applied Physics}, 104(4):044302, 2008.

\bibitem{venezian1994}
G.~Venezian.
\newblock {On the resistance between two points on a grid}.
\newblock {\em American Journal of Physics}, 62(11):1000, 1994.

\bibitem{atkinson1999}
D.~Atkinson and F.J.~Van Steenwijk.
\newblock {Inﬁnite resistive lattices}.
\newblock {\em American Journal of Physics}, 67(6):486--492, 1999.

\bibitem{cserti2000}
J.~Cserti.
\newblock {Application of the lattice Green’s function for calculating the
  resistance of an infinite network of resistors}.
\newblock {\em American Journal of Physics}, 68(10):896, 2000.

\bibitem{cserti2002}
J.~Cserti, G.~Dávid, and A.~Piróth.
\newblock {Perturbation of infinite networks of resistors}.
\newblock {\em American Journal of Physics}, 70(2):153, July 2002.

\bibitem{Stauffer1992}
D.~Stauffer and A.~Aharony.
\newblock {\em {Introduction to percolation theory}}.
\newblock Taylor \& Francis, 1992.

\bibitem{Pike1974}
G.~Pike and C.~Seager.
\newblock {Percolation and conductivity: A computer study. I}.
\newblock {\em Physical Review B}, 10(4):1421--1434, August 1974.

\bibitem{Seager1974}
C.~Seager and G.~Pike.
\newblock {Percolation and conductivity: A computer study. II}.
\newblock {\em Physical Review B}, 10(4):1435--1446, August 1974.

\bibitem{Munson-McGee1991}
S.~Munson-McGee.
\newblock {Estimation of the critical concentration in an anisotropic
  percolation network}.
\newblock {\em Physical Review B}, 43(4):3331--3336, February 1991.

\bibitem{Celzard1996}
A.~Celzard, E.~McRae, C.~Deleuze, M.~Dufort, G.~Furdin, and J.F.
  Mar\^{e}ch\'{e}.
\newblock {Critical concentration in percolating systems containing a
  high-aspect-ratio filler}.
\newblock {\em Physical Review B}, 53(10):6209--6214, March 1996.

\bibitem{Fuhrer2000}
M.S. Fuhrer, J.~Nygard, L.~Shih, M.~Forero, Y.G. Yoon, M.S.C. Mazzoni, H.J.
  Choi, J.~Ihm, S.G. Louie, A.~Zettl, and P.L. McEuen.
\newblock {Crossed nanotube junctions}.
\newblock {\em Science}, 288:494--497, April 2000.

\bibitem{Nirmalraj2009}
P.N. Nirmalraj, P.E. Lyons, S.~De, J.N. Coleman, and J.J. Boland.
\newblock {Electrical connectivity in single-walled carbon nanotube networks.}
\newblock {\em Nano letters}, 9(11):3890, November 2009.

\bibitem{Brandbyge2002}
M.~Brandbyge, J.-L. Mozos, P.~Ordej\'{o}n, J.~Taylor, and K.~Stokbro.
\newblock {Density-functional method for nonequilibrium electron transport}.
\newblock {\em Physical Review B}, 65(16):165401, March 2002.

\bibitem{Rocha2005}
A.R. Rocha, V.M. Garc\'{\i}a-Su\'{a}rez, S.W. Bailey, C.J. Lambert, J.~Ferrer,
  and S.~Sanvito.
\newblock {Towards molecular spintronics.}
\newblock {\em Nature materials}, 4(4):335, April 2005.

\bibitem{Yoon2001}
Y.G. Yoon, M.~Mazzoni, H.~Choi, J.~Ihm, and S.G. Louie.
\newblock {Structural Deformation and Intertube Conductance of Crossed Carbon
  Nanotube Junctions}.
\newblock {\em Physical Review Letters}, 86(4):688--691, January 2001.

\bibitem{Mowbray2009}
D.~Mowbray, C.~Morgan, and K.~Thygesen.
\newblock {Influence of O2 and N2 on the conductivity of carbon nanotube
  networks}.
\newblock {\em Physical Review B}, 79(19):195431, May 2009.

\bibitem{Caroli1971}
C.~Caroli, R.~Combescot, P.~Nozieres, and D.~Saint-James.
\newblock {Direct calculation of the tunnelling current}.
\newblock {\em Journal of Physics C: Solid State Physics}, 4:916, 1971.

\bibitem{Datta1997}
S.~Datta.
\newblock {\em {Electronic transport in mesoscopic systems}}.
\newblock Cambridge University Press, Cambridge, UK, 1997.

\bibitem{Mathon1997a}
J.~Mathon, A.~Umerski, and M.~Villeret.
\newblock {Oscillations with Co and Cu thickness of the
  current-perpendicular-to-plane giant magnetoresistance of a Co/Cu/Co (001)
  trilayer}.
\newblock {\em Physical Review B}, 55(21):14378, 1997.

\bibitem{Mathon1997b}
J.~Mathon.
\newblock {Tight-binding theory of tunneling giant magnetoresistance}.
\newblock {\em Physical Review B}, 56(18):11810, 1997.

\bibitem{Doherty2009}
E.M. Doherty, S.~De, P.E. Lyons, A.~Shmeliov, P.N. Nirmalraj, V.~Scardaci,
  J.~Joimel, W.J. Blau, J.J. Boland, and J.N. Coleman.
\newblock {The spatial uniformity and electromechanical stability of
  transparent, conductive films of single walled nanotubes}.
\newblock {\em Carbon}, 47(10):2466--2473, August 2009.

\bibitem{Pereira2009}
L.F.C. Pereira, C.G. Rocha, A.~Latge, J.N. Coleman, and M.S. Ferreira.
\newblock {Upper bound for the conductivity of nanotube networks}.
\newblock {\em Applied Physics Letters}, 95(12):123106, 2009.

\bibitem{Segal2009}
Michael Segal.
\newblock {Nanotubes: Effective networking}.
\newblock {\em Nature Nanotechnology}, (doi:10.1038/nnano.2009.316), October
  2009.

\bibitem{Hecht2006}
D.~S. Hecht, L.~Hu, and G.~Gruner.
\newblock {Conductivity scaling with bundle length and diameter in single
  walled carbon nanotube networks}.
\newblock {\em Applied Physics Letters}, 89(13):133112, 2006.

\bibitem{Shin2009}
D.H. Shin, H.C. Shim, J.-W. Song, S.~Kim, and C.-S. Han.
\newblock {Conductivity of films made from single-walled carbon nanotubes in
  terms of bundle diameter}.
\newblock {\em Scripta Materialia}, 60(8):607--610, April 2009.

\bibitem{Economou2006}
E.N. Economou.
\newblock {\em {Green's functions in quantum physics}}.
\newblock Springer, Berlin, third edition, 2006.

\bibitem{Geng2007}
H.-Z. Geng, K.K. Kim, K.P. So, Y.S. Lee, Y.~Chang, and Y.H. Lee.
\newblock {Effect of acid treatment on carbon nanotube-based flexible
  transparent conducting films.}
\newblock {\em Journal of the American Chemical Society}, 129(25):7758, June
  2007.

\end{thebibliography}

\end{document}